\documentclass[aps,prl,reprint,groupedaddress,showpacs,showkeys]{revtex4-1}

\usepackage{bm}
\usepackage{epsfig}
\usepackage{graphicx,graphics}
\usepackage{amsmath}
\usepackage{amsfonts}
\usepackage{amssymb}
\usepackage{epstopdf}

\begin{document}
	
	
	\title{ Membrane Stretching
		Elasticity of Lipid Vesicle and Thermal Shape Fluctuations}
	
	
	\author{Isak Bivas}
	\email[]{bivas@issp.bas.bg}
	\author{Nicholay S. Tonchev}
	\email[]{tonchev@issp.bas.bg}
	\affiliation{Institute of Solid State Physics, Bulgarian Academy of Sciences,
		72~Tzarigradsko chaussee blvd., Sofia~1784, Bulgaria}
	
	
	\date{\today}

	\begin{abstract}
		One of the most widely used methods for determination of the bending elasticity
		modulus  of model lipid membranes is the analysis of the shape fluctuations of
		nearly spherical lipid vesicles. The theoretical basis of this analysis is
		given by Milner and Safran. In their theory the stretching effects are not
		considered. In the present study we generalized their approach including the
		stretching effects  deduced after an application of  statistical
		mechanics of vesicles.
	\end{abstract}
	
	\pacs{87.16.dg, 87.16.dj, 82.70.Uv, 87.16.dm, 82.70.Kj}
	\keywords{lipid bilayer; modules of elasticity; vesicle; shape fluctuations}
	
	\maketitle
	
	\section{Introduction}
	The mechanical properties of biomenbranes determine to a great extent their
	structure and functioning. According to the model of Singer and
	Nicolson~\cite{Sin72}, the biomembrane consists of a lipid bilayer, in which
	integral proteins float. In the frames of this model, the mechanical properties
	of the biomembrane are tightly connected with those of the lipid bilayer.
	Because of its very small thickness compared to the square root of its area, the lipid bilayer
	may be appropriately modeled as two-dimensional flexible sheet and effectively
	studied theoretically, experimentally, and by computer simulations.
	
	The notion of membrane
	tension  defined  plays a key role in the consideration of the elastic deformation of the
	lipid  bilayers, bending and stretching.
	The importance of curvature energy  to the elasticity of  membrane has
	been emphasized by Helfrich in his pioneering paper from 1973~\cite{Helf73},
	while the stretching effects were mentioned only in passing.  A few years
	later
	Brochard, De Gennes  and Pfeuty~\cite{Bgp76} considered the vesicle membrane as
	compressible and revealed the role of the changes of the local density on the
	elastic energy.
	Due to some ambiguity around the definition of the
	surface tension, the study of this issue is continuously growing
	\cite{Mil87,Fauc89,Sei95,Saf99,Fap01,Hi04,Bif10,F11,Lli11,Nag12,S13,L14,Dim14}.
	In these papers, a number of questions  were raised. They concern: the
	physical meaning of the various definitions of tensions; the  consistency between
	reasonable approximations and theoretical predictions; equivalence of ensembles, which requires implicitly the existence of  thermodynamic limits, regimes where
	equilibrium thermodynamics  is far from being justified, etc. These phenomena are still a source of
	both theoretical and experimental challenges, see \cite{S13,L14,Hnf16} and
	refs.\ therein.
	
	The state of a closed bilayer (e.\ g.\ vesicle), that does not contain membrane reservoirs, and of an incompressible membrane
	conformation is determined by the  bending elasticity and constraints on the total
	surface area and on the enclosed volume. The vesicle has constrained area and constrained volume, since the number of the lipid molecule in the membrane is fixed.
	While it is technically  easy to implement the second constraint, it is difficult to handle the first one. Instead of working with fixed area, an effective tension as Lagrange
	multiplier conjugated to the fixed  membrane area has been
	used~\cite{Mil87,Sei95}, expecting that the two ensembles are equivalent.
	
	The aim of our study is to elucidate
	the theoretical basis of the analysis, initiated  by Milner and
	Safran~\cite{Mil87} in order to take into consideration the role of the
	membrane compression and tension in terms of stretching elasticity, which are
	introduced into the theory by nonlocal (anharmonic) terms.
	
	We restrict our considerations to mechanical systems and present
	some
	general characteristics of the mean-field approximation in the context of the
	statistical mechanics of vesicles.
	
	\section{The total deformation energy of the vesicle membrane}
	
	We considered lipid bilayers in liquid crystalline state, in which they can be modeled by
	two-dimensional liquids.
	Following Helfrich~\cite{Helf73}, we consider a small patch of the lipid bilayer
	with area $\Delta s$, tension $\sigma$, and area in its flat tension-free
	state $s_0$. Let $c_1$ and $c_2$ be the main curvatures of the patch
	under consideration. If the patch is tension-free, then the bending energy
	density
	$g_c(c_1,c_2)$ can be written in the following form:
	\begin{eqnarray}
	g_c(c_1,c_2)=\frac{1}{2}K_c(c_1+c_2-c_0)^2+\overline{K}_cc_1c_2, \label{h10}
	\end{eqnarray}
	where $K_c$ is the bending elasticity of the bilayer, $c_0$ is the spontaneous
	curvature of the membrane, and $\overline{K}_c$ is the saddle splay bending
	elasticity. Only symmetrical membranes (with $c_0=0$) will be considered later on.
	The bending energy $\Delta G_c$ of the  patch is:
	\begin{equation}
	\Delta G_c = g_c(c_1,c_2) s_0. \label{h13a}
	\end{equation}
	If the patch is not tension-free, then its stretching energy density $g_s$ is
	expressed via its tension $\sigma$ as:
	\begin{equation}
	g_s(\sigma)=\frac{1}{2}\frac{\sigma^2}{K_s}. \label{h15}
	\end{equation}
	 We assume that $\sigma$ is a constant all over the membrane (see~\cite{Biv02}). Then $\sigma$ can be expressed as:
	\begin{equation}
	\sigma= K_s \frac{\Delta s-s_0}{s_0},  \label{h15b}
	\end{equation}
	where $K_s$ is the stretching elasticity of the bilayer.
	The stretching energy $\Delta G_s$ and the total mechanical energy $\Delta G$ of
	the patch are:
	\begin{equation}
	\Delta G_s =g_s(\sigma) s_0. \label{h15a}
	\end{equation}
	and
	\begin{equation}
	\Delta G=\Delta G_c+\Delta G_s, \label{h16}
	\end{equation}
	respectively.
	
	Let us consider a vesicle whose membrane is a lipid bilayer. The shape
	fluctuations of the vesicle deform this bilayer and alter its mechanical energy.
	The total deformation energy $G(t)$ of the vesicle membrane is obtained by
	integration of $\Delta G(t)$ on the vesicle surface $S(t)$:
	\begin{equation}
	G(t)=\oint_{S(t)}\Delta G(t). \label{h16a}
	\end{equation}
	
	\section{Quasi-spherical vesicles: the Milner and Safran approach}
	We  consider a nearly spherical lipid vesicle with fixed (i.e. not fluctuating)
	volume $V$.
	Let $R=(\frac{3V}{4\pi})^{1/3}$ be the radius of a sphere with the
	same volume  $V$.  Let $R(\theta,\varphi)$ is the modulus of the
	radius-vector  of a point on the surface of the vesicle with polar coordinates
	$(\theta,\varphi)$ in laboratory reference frame with origin $0$  placed
	inside the vesicle. The dimensionless quantity $u(\theta,\varphi,t)$ is
	defined by the equation:
	\begin{equation}
	u(\theta,\varphi,t)=\frac{R(\theta,\varphi,t)-R}{R}, \label{e5}
	\end{equation}
	where $t$ is the time variable. The function $u(\theta, \varphi,t)$ is
	decomposed
	in a series of spherical harmonics as follows:
	\begin{equation}
	u(\theta,\varphi,t) =
	\sum_{n=0}^{n_{max}}\sum_{m=-n}^{n}
	u_n^m(t) Y_n^m(\theta,\varphi), \label{e6}
	\end{equation}
	where $Y_n^m(\theta,\varphi)$ is the orthonormal basis (for simplicity chosen
	real) of the spherical harmonics functions. A cut-off $n_{max}\sim
	R/\lambda$ is introduced in the sum, where $\lambda$ is of the order of the
	intermolecular distance. As the harmonics with indexes $n=1$ and $m=-1,0,1$
	correspond to pure translation of the vesicle, the origin $O$ can be chosen in a
	way that $u_1^m=0$.
		
	As it has been shown by Milner and Safran \cite{Mil87}, because of the requirement for volume conservation the
	amplitude $u_0^0(t)$ can be expressed
	as:
	\begin{equation}
	u_0^0(t)=-\frac{1}{2\sqrt{\pi}}\sum_{n=2}^{n_{max}}\sum_{m=-n}^{n}
	[u_n^m(t)]^2. \label{e7}
	\end{equation}
	
	Then the
	mechanical energy $G(t)$ of the vesicle, obtained after the integration in the
	rhs of Eq.~\eqref{h16a}, is:
	\begin{eqnarray}
	G(t)&=&\frac{1}{2}K_c
	\sum_{n=2}^{n_{max}}\sum_{m=-n}^{n}\{ (n-1)(n+2)\nonumber \\
	&\times& [n(n+1)+ \overline{\sigma}][u_n^m(t)]^2\}, \label{h19}
	\end{eqnarray}
	where the quantity $\overline{\sigma}$ is considered as a Lagrange multiplier,
	not fluctuating with time, which ensures the mean area of the vesicle membrane
	to
	be equal to some prescribed value. In Eq.~\eqref{h19} the following
	dimensionless expression for $\overline{\sigma}$:
	\begin{equation}
	\overline{\sigma} = \frac{R^2}{K_c} \sigma, \label{h20}
	\end{equation}
	where $\sigma$ is the tension of the membrane,  has been introduced. The
	contribution of the saddle-splay elasticity has been disregarded  due to the Gauss-Bonnet theorem
which assures that if the topology of
	the vesicle does not change, then the contribution of this elasticity does not
	depend on the shape fluctuations of the vesicle. The energy $G(t)$ from
	Eq.~\eqref{h19} can be considered as sum of the energies of not interacting
	"oscillators". It is worth noting that this energy does not depend on the
	stretching elasticity $K_s$ of the membrane and is a function  only of $K_c$,
	$R$, and $\sigma$.
	The expression $G(t)$ in
	Eq.~\eqref{h19} can be implicitly considered as a Hamiltonian of a system
	consisting of not interacting oscillators, whose amplitudes $u_n^m(t)$ are the
	dynamical variables of the system. As a result, for $n\ge 2$ the time mean
	squares $\langle [u_n^m(t)]^2 \rangle$ of the amplitudes $u_n^m(t)$ of the
	different fluctuation modes are:
	\begin{equation}
	\langle [u_n^m(t)]^2 \rangle = \frac{kT}{K_c}
	\frac{1}{(n-1)(n+2)[n(n+1)+\overline{\sigma}]} \label{h21}
	\end{equation}
	where $kT$ is the Boltzmann factor. Namely this result is used for the analysis
	of the shape fluctuation of nearly spherical vesicles.
	One can conclude that Milner and Safran \cite{Mil87} developed a theory  with membrane
	possessing bending elasticity, having no stretching elasticity and possessing
	different from zero tension, which does not depend on the shape fluctuations of
	the vesicle.
	

	\section{Model-dependent derivations}
	
	Later on we consider a model Hamiltonian of the fluctuating
	vesicle, whose membrane has stretching and bending elasticity as well.
	Manipulating this Hamiltonian in an appropriate way, i.\ e.\ making a mean field
	approximation (see below), we shall effectively  approximate the fluctuating
	vesicle by a system in form
	of not interacting harmonic oscillators.
	
	\subsection{The surface tension}

	Let us denote by $H(U)= H(u_2^{-2},
	u_2^{-1},\dots,u_{n_{max}}^{n_{max}})$   the model
	Hamiltonian describing a fluctuating vesicle. The symbol  $U$ is used as
	shorthand for the real value  functions $(u_2^{-2}, u_2^{-1},\dots,
	u_{n_{max}}^{n_{max}})$ which are the spherical harmonics amplitudes, appearing
	in
	the the expansion of the vesicle shape fluctuations from the equivalent volume
	sphere with radius $R$ (see Eq.~\eqref{e6}). Further on, the notation $u_n^m$ is
	used as an abbreviation of $u_n^m$(t).
	
	One can present the area functional of the membrane $S(U)$ in the
	form (see~\cite{Mil87,Fauc89,Hi04}):
	\begin{equation}
	S(U)=4\pi R^2 +\Delta S(U), \label{r18}
	\end{equation}
	where the quantity $\Delta S(U)$ is the excess area of the vesicle (the
	difference between the area of the vesicle's membrane and the area $4\pi R^2$
	of a sphere with a volume equal to that of the vesicle).

	 In terms
		of  $u_n^m$ the excess area of the vesicle $\Delta S(U)$ is presented in the form  ~\cite{Mil87,Sei95}:
	\begin{equation}
	\Delta S(U)=  \frac{R^2}{2}
	\Bigg[ \sum_{n=2}^{n_{max}}\sum_{m=-n}^n
	(n-1)(n+2) (u_n^m)^2 \Bigg]. \label{e18}
	\end{equation}
	
	Note that in Eq.\eqref{e18} the requirement for volume conservation  has been used in order to
	exclude the amplitude $u_0^0$.
	Evidently, $\Delta S(U) \ge 0$. This property of $\Delta S(U)$ does not depend on the form of the Hamiltonian.
	
	By definition the
	vesicle tension $\sigma(U)$ is given by the expression:
	\begin{equation}
	\sigma(U) = K_s\frac{S(U)-S_0}{S_0}, \label{r17}
	\end{equation}
	where $S_0$ is the mean area of the vesicle membrane in its flat tension
	free state

	It will be shown that the experimentally measurable quantity is proportional to:
	\begin{equation}
	\langle \sigma(U)\rangle_{H(U)} = K_s\frac{\langle S(U)\rangle_{H(U)}-S_0}{S_0}.
	\label{r17a}
	\end{equation}
	With the symbol $\langle A  \rangle_{H(U)}$ we denote the thermodynamic
	average of some quantity $A$ calculated with the Hamiltonian ${H(U)}$:
	\begin{equation}
	\langle A \rangle_{H(U)}=\left\{Z[{H(U)}]\right\}^{-1}\int d{U}
	A\exp{\left[-\frac{{H(U)})}{kT}\right]},
	\end{equation}
	where
	\begin{equation}
	Z[{H(U)}] =\int d{U} \exp{\left[-\frac{{H(U)}}{kT}\right]}
	\label{e22b}
	\end{equation}
	is the statistical sum of the model.
	Indeed, one can  consider  $\langle \sigma(U)\rangle_{H(U)}$ as the  mechanical tension
	and Eq.\eqref{r17a} as the definition of the area compressibility modulus $K_s$,
	when the mean area of the vesicle membrane $\langle S(U)\rangle_{H(U)}$ deviates from the mean area of its flat tension-free state $S_0$.
	
	\subsection{Model Hamiltonian and Bogoliubov variational inequalities}
	
	Let us consider a  model system, having model Hamiltonian, which in the
	general case is complicated enough and does not allow to find analytical
	expressions for the investigated thermodynamic quantities. We will replace the model system with
	a trial one (named approximating), consisting of independent (not interacting) oscillators.
 We call mean-field approximation the replacement of the model system with the approximating one.
	Clearly, such a definition does not determine the  approximating Hamiltonian in
	an unequivocal way. It can be expected that the mean-field approximation will be a
	good approximation when the approximating Hamiltonian is close in some sense to the model one.
	A natural
	question arises how to ensure the closeness between these Hamiltonians. A common
	and well-known approach in statistical mechanics is to use the Bogoliubov
	variational inequalities and the approximating
	Hamiltonian method (presented below) for the construction  of an approximating
	system,
	whose free energy is close to the free energy of the model system. Namely such
	an
	approach will be used for the implementation of the approximating system.
	
	The Hamiltonian
	under consideration is presented in the following form:
	\begin{equation}
	H(U)=H_c(U)+H_s(U), \label{IB}
	\end{equation}
	where
	\begin{equation}
	H_c(U)=  \frac{1}{2}K_c \sum_{n=2}^{n_{max}}\sum_{m=-n}^n
	(n-1)n(n+1)(n+2)(u_n^m)^2    \label{e17}
	\end{equation}
	is the   bending energy of the vesicle   with quasi-spherical geometry, and
	\begin{equation}
	H_s(U)=\frac{1}{2}\frac{ [\sigma(U)]^2}{K_s}S_0 \label{Hs1}
	\end{equation}
	is  its  corresponding stretching energy, expressed via the membrane vesicle
	tension
	$\sigma(U)$. The membrane tension $\sigma(U)$  is again assumed to be the same
	all over the membrane.

	The term  $H_s(U)$  is nonlinear with respect to the squares
	 of the amplitudes $u_n^m$.  To overcome  this
	obstacle  we linearize the Hamiltonian \eqref{Hs1}
	using the Bogoliubov  variational inequalities (for historical remarks and a
	list of different applications see, e.g. \cite{Bdt00}).
	The Bogoliubov variational inequalities are rigorous relations between the free
	energy  $f[H]$ of a valid Hamiltonian $H$ and free energy density
	$f[H_{app}(X)]$
	of a presumably more simple Hamiltonian $H_{app} (X)$, depending on a variational parameter $X$. In
	their most convenient form the inequalities are given by
	\begin{eqnarray}
	&&\langle  H-H_{app}(X) \rangle_{H}  \nonumber \\
	&& \le  f[H]-f[H_{app}(X)] \le  \langle H-H_{app}(X) \rangle_{H_{app}(X)},
	\label{e46}
	\end{eqnarray}
	where $\langle H-H_{app}(X) \rangle_{H}$ is the thermodynamic average of the
	quantity $H-H_{app}(X)$, calculated with the Hamiltonian $H$,  $\langle
	H-H_{app}(X) \rangle_{H_{app}(X)}$ is the thermodynamic average of the same
	quantity calculated with the Hamiltonian $H_{app}(X)$, $Z[H]$ and $f[H]=-kT\ln\{
	Z[H]\}$ are the partition function and the free energy of a system with
	Hamiltonian
	$H$, $Z[H_{app}(X)]$ and $f[H_{app}(X)]=-kT\ln\{ Z[H_{app}(X)]\}$ are the
	partition function and the free energy of a system with Hamiltonian
	$H_{app}(X)$.
	The variational parameter $X$
	must be determined from the condition of the best approximation.
	
	The use  of these inequalities in the  statistical mechanics of a lipid vesicle
	has
	been announced  in \cite{Biv14}.
	
	\subsection{The approximating Hamiltonian}
	
	The model Hamiltonian
	$H(U)$  has been rewritten in  an alternative form  that makes the application
	of the Bogoliubov
	variational inequalities (see \eqref{e46}) more convenient:
	\begin{equation}
	H(U)={\cal T}(U)+[{\cal A}(U)]^2, \label{e141}
	\end{equation}
	where
	\begin{equation}
	{\cal T}(U)=\frac{1}{2} K_c
	\sum_{n=2}^{n_{max}}\sum_{m=-n}^n(n-1)(n+2)[n(n+1)+\overline{\sigma}_0](u_n^m)^2.
	\label{e140}
	\end{equation}
	Here and further on the bare over the quantity means dimensionless due to the multiplier $R^2/K_c$
	\begin{equation}
	\overline{\sigma}_0=\frac{R^2}{K_c}\sigma_0
	\label{e16}
	\end{equation}
	\begin{equation}
	\sigma_0=K_s\frac{4\pi R^2-S_0}{S_0}, \label{e166}
	\end{equation}
	and
	\begin{equation}
	{\cal A}(U) =\sqrt{\frac{K_s}{2S_0}}\Delta S(U). \label{e39}
	\end{equation}
	Evidently the following identity holds:
	\begin{equation}
	[{\cal A}(U)]^2 = 2X{\cal A}(U)-X^2+[{\cal A}(U)-X]^2, \label{e142}
	\end{equation}
	where $X$ is an arbitrary real number. We define the linearized Hamiltonian
	$H_{app}(U,X)$ as:
	\begin{equation}
	H_{app}(U,X)= {\cal T}(U) +2X{\cal A}(U)-X^2 \label{e44}
	\end{equation}
	The last equation is obtained from Eq.~\eqref{e141} by removing the term $[{\cal A}(U)-X]^2$ from the right-hand-side. The defined in
	this way Hamiltonian $H_{app}(U,X)$ is linear with respect to
	$(u_n^m)^2$.
	
	From Eqs.~\eqref{e140}, \eqref{e39}, and \eqref{e44} we obtain:
	\begin{equation}
	H_{app}(U,X)=\sum_{n=2}^{n_{max}}\sum_{m=-n}^{n}a_n(X) (u_n^m)^2 -X^2,
	\label{AH1}
	\end{equation}
	where
	\begin{equation}
	a_n(X)=\frac{1}{2}K_c(n-1)(n+2)[n(n+1)+\overline{\Sigma}_{app}(X)],
	\label{AH2}
	\end{equation}
	and
	\begin{equation}
	\overline{\Sigma}_{app}(X)= \overline{\sigma}_0 + \overline{\sigma}_1 X \label{e58}
	\end{equation}
	with $\overline{\sigma}_0$ from Eq.~\eqref{e16}, and
	\begin{equation}
	\overline{\sigma}_1= \frac{R^2}{K_c}\sqrt{\frac{2 K_s}{S_0}}.
	\label{e59}
	\end{equation}
	Since the thermal average of a nonnegative quantity is nonnegative it follows that:
	\begin{equation}
	\langle H(U)-H_{app}(U,X) \rangle_{H(U)} =\langle [{\cal A}(U)-X]^2
	\rangle_{H(U)} \ge 0. \label{e147}
	\end{equation}
	Then Eqs.~\eqref{e46} and \eqref{e147} imply that for each $X$:
	\begin{equation}
	0 \le f[H(U)] - f[H_{app}(U,X)] \le \langle [{\cal A}(U)-X]^2
	\rangle_{H_{app}(U,X)}. \label{e148}
	\end{equation}
	We define $\widetilde{X}$ as the solution of the equation:
	\begin{equation}
 \frac{\partial f[H_{app}(U,X)]}{\partial X}=0
	\label{e51}
	\end{equation}
It can be shown that the equation above has only one solution, namely $\widetilde{X}$ and
$\widetilde{X}$ satisfies the condition:
\begin{equation}
	f[H_{app}(U,\widetilde{X})]=\max_X f[H_{app}(U,X)]. \label{e149}
	\end{equation}
From Eq.~\eqref{e148} it follows that:
	\begin{equation}
	f[H_{app}(U,\widetilde{X})]\le f[H(U)].
	\end{equation}
	Consequently, the free energy $f[H_{app}(U,\widetilde{X})]$ of the ensemble of
	not
	interacting oscillators is the best approximation from below of the free energy,
	corresponding to the model Hamiltonian $H(U)$.
	
	\subsection{The self-consistent equation}
	
	To present Eq.~\eqref{e51} in explicit form we start from the Helmholtz free
	energy $f[H_{app}(U,X)]$:
	\begin{equation}
	f[H_{app}(U,X)]=-kT\ln\{Z[H_{app}(U,X)]\}, \label{e60}
	\end{equation}
	where $Z[H_{app}(U,X)]$ is the partition function of the approximating system:
	\begin{equation}
	Z[H_{app}(U,X)]  =\int dU
	\Bigg\{\exp{\Bigg[-\frac{H_{app}(U,X)}{kT}\Bigg]}\Bigg\}. \label{e61}
	\end{equation}
	As a result we obtain:
	\begin{eqnarray}
	&& f[H_{app}(U,X)]  =  -kT \nonumber \\
	&\times &\ln{\left\{\int dU \exp \left[-\frac{{\cal T}(U)+2X{\cal
				A}(U)}{kT}\right]\right\}} -X^2, \label{fr1)}
	\end{eqnarray}
	and:
	\begin{eqnarray}
	&&\frac{\partial f[H_{app}(U,X)]}{\partial X} \nonumber   \\
	&&=2[\langle {\cal A}(U) \rangle_{H_{app}(U,X)} -X] =0.
	\label{e52}
	\end{eqnarray}
	Consequently, Eq.~\eqref{e51} can be written in the following equivalent form:
	\begin{equation}
	\langle {\cal A}(U) \rangle_{H_{app}(U,X)} -X =0. \label{e444}
	\end{equation}
	This is a typical self-consistent equation for the parameter $X$.
	
	After simple calculations,
	from Eqs.~\eqref{AH1} and \eqref{e61} we obtain
	for the free energy $f[H_{app}(U,\widetilde{X})]$ of the approximating system:
	\begin{eqnarray}
	&&f[H_{app}(U,\widetilde{X})] \nonumber \\
	&&= kT \sum_{n=2}^{n_{max}}
	\frac{2n+1}{2} \ln\left\{(n-1)(n+2)[n(n+1)+ \overline{\Sigma}_{app}(\widetilde{X})] \right\}
	\nonumber \\
	&&\mbox{} -\left(\widetilde{X} \right)^2  + kT\frac{N}{2}\ln\left(\frac{K_c}{2\pi
		kT}\right). \label{e62}
	\end{eqnarray}
	In the equation above $N\approx(n_{max})^2$ is the number of lipid molecules in
	the vesicle membrane.

	Further on, in order to be unambiguous  we will use  notations linearized and
	approximating Hamiltonian
	for $H_{app}(U,X)$ and
	$H_{app}(U,\widetilde{X})$, respectively.
	
	The mean squares $\langle (u_n^m)^2\rangle_{H_{app}(U,X)}$, calculated
	by the linearized Hamiltonian $H_{app}(U,X)$, are:
	\begin{eqnarray}
	&&\langle (u_n^m)^2\rangle_{H_{app}(U,X)}\nonumber\\
	&&= \frac{k T}{K_c}\frac{1}{(n-1)(n+2)[n(n+1)+\overline{\Sigma}_{app}(X)]}.
	\label{e67}
	\end{eqnarray}
	Except for difference in the interpretation of the last term
	in the denominator of \eqref{e67} the result for $\langle (u_n^m)^2\rangle_{H_{app}(U,X)}$ has been obtained by many authors
	\cite{Mil87,Sei95,Bif10}.
	
	As it can be seen after comparison with Eq.~\eqref{h20}, the
	result for $\langle (u_n^m)^2\rangle_{H_{app}(U,X)}$ formally reproduces the result of Milner and Safran. The difference between our and their result consists in the interpretation. In their theory  $\overline{\sigma}$ is considered as a
	Lagrange multiplier, while in our theory $\overline{\Sigma}_{app}(X)$ with $X=\widetilde{X}$ is obtained by application of the principles of the statistical mechanics to  concrete model Hamiltonian $H(U)$.

Let us for convenience  introduce  the shorthand
	\begin{equation}
	\overline{\Sigma}_{app} = \overline{\Sigma}_{app}(\widetilde{X}).
	\label{53}
	\end{equation}
Then $\overline{\Sigma}_{app}$ can be obtained from the self-consistent equation \eqref{e444} as follows.
	With the help of Eqs.~\eqref{e39}, \eqref{e59}, and \eqref{e67},   Eq.
	\eqref{e444} can be rewritten as:
	\begin{equation}
	X=\frac{kT\overline{\sigma}_1}{4}\sum_{n=2}^{n_{max}} \frac{2n+1}{n(n+1)+
		\overline{\Sigma}_{app}(X)}. \label{e68}
	\end{equation}
	By multiplying the two sides of Eq.~\eqref{e68} with
	$\overline{\sigma}_1$ (see Eq.~\eqref{e59}) and adding to them
	$\overline{\sigma}_0$ (see Eq.~\eqref{e16}), we get the following form the self-consistent equation
	\eqref{e444},
	written in a different way:
	\begin{equation}
	\overline{\Sigma}_{app} =\overline{\sigma}_0+
	\frac{kT}{2}\frac{K_s}{S_0}\frac{R^4}{(K_c)^2}
	\sum_{n=2}^{n_{max}} \frac{2n+1}{n(n+1)+\overline{\Sigma}_{app}}.
	\label{e68c}
	\end{equation}
	
	The dependence of $\langle (u_n^m)^2\rangle_{H_{app}(U,\widetilde{X})}$ on $K_s$ is hidden
	in
	$\overline{\Sigma}_{app}$ (see Eqs.~\eqref{e67} and \eqref{e68c}).

	Inserting Eq.~\eqref{e16} into Eq.~\eqref{e68c},  we obtain:
	\begin{eqnarray}
	&&\overline{\Sigma}_{app}=\frac{R^2}{K_c}\frac{K_s}{S_0}
	\bigg\{4\pi R^2 \nonumber \\
	&&\mbox{}+ \frac{kT}{K_c}\frac{R^2}{2}
	\sum_{n=2}^{n_{max}} \frac{2n+1}{n(n+1)+\overline{\Sigma}_{app}}-S_0 \bigg\}.
	\label{e68d}
	\end{eqnarray}

	 We introduce the new quantity $\overline{\Sigma}_{app}^0$  through the relation:
		\begin{equation}
		S_0=4\pi R^2
		+\frac{kT}{K_c}\frac{R^2}{2}\sum_{n=2}^{n_{max}}\frac{2n+1}{n(n+1)+
			\overline{\Sigma}_{app}^0}. \label{bt1}
		\end{equation}
		Its physical meaning will be revealed later (see the comment concerning  Eq.~\eqref{bt4}.
		
\begin{figure}[t]
			\includegraphics[width=3in,height=3in]{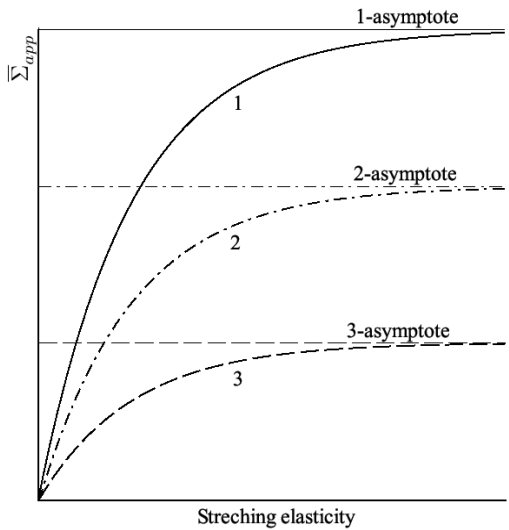}
			\caption{\label{fig11} Schematic representation of the dependence of the vesicle
				tension $\overline{\Sigma}_{app}$ (see Eq.~\eqref{e68d}) on its stretching
				elasticity $K_s$. The three curves 1, 2, and 3 refer to three vesicles with identical radiuses
				$R$, bending elasticities $K_c$, and temperature $T$, and with different
				tension-free areas $S_0$ of their membranes. $S_0$ of curve 1 is less than that of curve 2, which is less than this of curve 3. When $K_s$ tends to $\infty$, the dependences of this kind tends to  horizontal asymptotes: $\overline{\Sigma}_{app}=\overline{\Sigma}_{app}^0$(see below). In the present case the asymptotes, corresponding to the curves 1, 2, and 3, are denoted as 1-asymptote, 2-asymptote, and 3-asymptote, respectively.}
		\end{figure}

		Using Eq.~\eqref{bt1}, Eq.~\eqref{e68d} can be presented in the
		following form:
		\begin{eqnarray}
		\overline{\Sigma}_{app}&=&\frac{R^4}{2K_c}\frac{K_s}{S_0}
		\frac{kT}{K_c}\nonumber \\
		&\times& \sum_{n=2}^{n_{max}}
		\frac{(2n+1)(\overline{\Sigma}_{app}^0-\overline{\Sigma}_{app})}{[n(n+1)+\overline{\Sigma}_{app}]
		[n(n+1)+\overline{\Sigma}_{app}^0]}.
		\label{bt2}
		\end{eqnarray}
	Eq.~\eqref{bt2} shows the dependence $\overline{\Sigma}_{app}(\frac{K_s}{K_s})$ for different values of $\overline{\Sigma}_{app}^0$ (respectively  $S_0$) at fixed $K_c$, $R$, and $T$.
For three different values of $S_0$, $\overline{\Sigma}_{app}$ is schematically presented on Fig.~\eqref{fig11}. The obtained results allow  to determine   the values of  $\overline{\Sigma}_{app}(K_s)$ when $K_s\rightarrow 0$ and $K_s \rightarrow \infty$. For the first limit the result is:
\begin{equation}
\lim_{K_s\rightarrow 0}\overline{\Sigma}_{app}(K_s)=0. \label{bt3}
\end{equation}
 The analogous case in the Milner and Safran approach is the case when $\overline{\sigma}=0$ (see Eqs.~\eqref{h19} and \eqref{h20}).

The second limit is exactly the auxiliary tension $\overline{\Sigma}_{app}^0$:
\begin{equation}
\lim_{K_s\rightarrow \infty}\overline{\Sigma}_{app}(K_s)= \overline{\Sigma}_{app}^0. \label{bt4}
\end{equation}
  This case has no analogue  in the theory of Milner and Safran.

We note, that by taking these limits the relation $K_c \sim K_s\times d^2$ where $d$ is the thickness of the membrane \cite{Helf73} no longer holds. Our aim is to check the behavior of the vesicle when the often used approximations of not stretchable or infinitely stretchable membrane. Evidently the above relation cannot be kept at such approximations. Our results show that in both cases the tension of the membrane is finite and  well-defined.

	From Eqs.~\eqref{e18} and \eqref{e67} it follows that:
	\begin{eqnarray}
	&&\langle \Delta S(U)\rangle_{H_{app}(U,\widetilde{X})}    =
	\frac{kT}{K_c}\frac{R^2}{2} \sum_{n=2}^{n_{max}}
	\frac{2n+1}{n(n+1)+\overline{\Sigma}_{app}}.\ \ \ \ \  \label{e78}
	\end{eqnarray}
	With the help of Eqs.~\eqref{r18}, \eqref{e18}, \eqref{e67}, and \eqref{e78},
	Eq.~\eqref{e68d} becomes:
	\begin{eqnarray}
	\overline{\Sigma}_{app}=\frac{R^2}{K_c}&&K_s\frac{\langle
	S(U)\rangle_{H_{app}(U,\widetilde{X})}-S_0}{S_0}\nonumber \\
	&&=\frac{R^2}{K_c}
	\langle \sigma(U) \rangle_{H_{app}(U,\widetilde{X})}, \label{ks1}
	\end{eqnarray}
	where $\sigma(U)$ is the true (not normalized) tension of the membrane (see
	Eqs.~\eqref{r17} and~\eqref{r17a}).
	
	\subsection{The stretching elasticity $K_s$ }
	
	One of the key points of this work is the elucidation of the role of the self-consistent equation. It is shown that this equation permits to express the stretching elasticity $K_s$ via experimentally accessible quantities.
	
	The self-consistent equation can be written in the following form:
	\begin{equation}
	K_s=\frac{\overline{\Sigma}_{app}\frac{S_0K_c}{R^2}}
	{\bigg[ \frac{kT}{K_c}\frac{R^2}{2}
		\sum_{n=2}^{n_{max}}
		\frac{(2n+1)(\overline{\Sigma}_{app}^0-\overline{\Sigma}_{app})}{[n(n+1)+\overline{\Sigma}_{app}]
			[n(n+1)+\overline{\Sigma}_{app}]} \bigg]}
	\label{e68e}
	\end{equation}
	It is clear that when the denominator of the fraction on the rhs of the above 	equation tends to zero, then $K_s \rightarrow \infty$.  The denominator is zero only if  $\overline{\Sigma}_{app}=\overline{\Sigma}_{app}^0$ (see the asymptotes on Fig.1).
	
	 In order to obtain experimentally $K_s$ it is sufficient to calculate the dependence of $\overline{\Sigma}_{app}$ on  $K_c$ and $K_s$ (i.e. to obtain from Eq.~\eqref{e68c} $\overline{\Sigma}_{app}(K_c,K_s)$) and after that to calculate the dependence of $\langle(u_n^m)^2\rangle_{H_{app}(U,\widetilde{X})}$ on $K_c$, $K_s$, and  $\overline{\Sigma}_{app}(K_c,K_s)$, (i.e. to obtain from Eq.~\eqref{e67}, with $X=\widetilde{X}$, $\langle(u_n^m)^2\rangle_{H_{app}(U,\widetilde{X})}[K_c,K_s, \overline{\Sigma}_{app}(K_c,K_s)]$).

	 In the general case $\overline{\Sigma}_{app}$ depends on $K_c$ (bending elasticity), $K_s$ (stretching elasticity), $R$ (the radius of the vesicle), and $S_0$ (the area of the vesicle membrane in its flat tension-free state). We  consider the case when the last two parameters can be determined or measured by independent methods and are not correlated with $K_c$ and $K_s$. Fitting $\langle(u_n^m)^2\rangle_{H_{app}(U,\widetilde{X})}$ with $K_c$ and $K_s$, we can determine the stretching elasticity of the membrane, deduced from the analysis of the thermal fluctuations of the vesicle shape.
	
		\begin{figure}[tbh]
			\includegraphics[width=3.6in,height=2.8in]{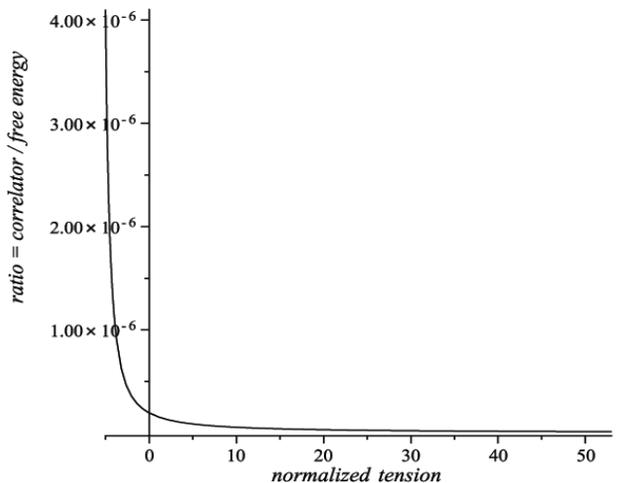}
			\caption{\label{fig2} The dependence of the ratio $correlator / free~energy$ on the normalized tension  $\overline{\Sigma}_{app}$ is shown. The numerical results are obtained with the given above numerical values of the quantities $K_s,K_c,R,S_0$.}
		\end{figure}

	\section{The closeness of the model Hamiltonian to the approximating Hamiltonian}
	
	We continue with the calculation of $\langle [{\cal A}(U)-\widetilde{X}]^2
	\rangle_{H_{app}(U,\widetilde{X})}$. Obviously, we have:
	\begin{eqnarray}
	&&\langle [{\cal A}(U)-\widetilde{X}]^2
	\rangle_{H_{app}(U,\widetilde{X},\sigma_0)} = \langle [{\cal A}(U)]^2
	\rangle_{H_{app}(U,\widetilde{X},\sigma_0)} \nonumber \\
	&& \mbox{}- 2 \langle
	{\cal A}(U) \rangle_{H_{app}(U,\widetilde{X},\sigma_0)}\widetilde{X}
	+\left(\widetilde{X}\right)^2
	\end{eqnarray}
	From Eqs.~\eqref{e39}, \eqref{e59}, and \eqref{e67} we obtain:
	\begin{eqnarray}
	\langle {\cal A}(U)\rangle_{H_{app}(U,\widetilde{X})}
	=\frac{kT\sigma_1}{4}\sum_{n=2}^{n_{max}}
	\frac{2n+1}{[n(n+1)+\overline{\Sigma}_{app}]}  \label{e72}
	\end{eqnarray}
	From Eqs.~\eqref{e444} it follows that:
	\begin{equation}
	\langle {\cal A}(U) \rangle_{H_{app}(U,\widetilde{X})}=\widetilde{X}. \label{e721}
	\end{equation}
	But from Eqs.~\eqref{e18} and \eqref{e39} it follows that:
	\begin{eqnarray}
	&&\langle [{\cal A}(U)]^2\rangle_{H_{app}(U,\widetilde{X})} =\frac{K_s}{2S_0}
	\frac{R^2}{4} \nonumber \\
	&&\times \sum_{n=2}^{n_{max}} \sum_{m=-n}^{n}
	\sum_{n'=2}^{n_{max}} \sum_{m'=-n'}^{n'} (n-1)(n+2) \nonumber \\ &&\times
	(n'-1)(n'+2) \langle(u_n^m)^2(u_{n'}^{m'})^2
	\rangle_{H_{app}(U,\widetilde{X})}.
	\end{eqnarray}
	Taking into account that the amplitudes $u_n^m$ are not correlated (the
	approximating Hamiltonian presents a system of not interacting oscillators)
and have a Gaussian distribution, we obtain that:
	\begin{equation}
	\langle (u_n^m)^4\rangle_{H_{app}(U,\widetilde{X})} =3[\langle
	(u_n^m)^2\rangle_{H_{app}(U,\widetilde{X})}]^2.
	\end{equation}
	After some tedious but simple calculations it follows that one gets
	\begin{eqnarray}
	&&\langle [{\cal A}(U)]^2\rangle_{H_{app}(U,\widetilde{X})}  \nonumber \\
	&& =[\langle {\cal A}(U)\rangle_{H_{app}(U,\widetilde{X})}]^2 \nonumber \\
	&& \mbox{}+\frac{K_s}{S_0}
	\frac{R^4}{4} \Bigg[\frac{kT}{K_c} \Bigg]^2 \sum_{n=2}^{n_{max}}
	\frac{2n+1}{[n(n+1)+\overline{\Sigma}_{app}]^2}.
	\end{eqnarray}
	Finally, we have:
	\begin{eqnarray}
	&&\langle [{\cal A}(U)-\widetilde{X}]^2
	\rangle_{H_{app}(U,\widetilde{X})} \nonumber \\
	&& = \frac{K_s}{S_0} \frac{R^4}{4}
	\Bigg[\frac{kT}{K_c} \Bigg]^2 \sum_{n=2}^{n_{max}}
	\frac{2n+1}{[n(n+1)+\overline{\Sigma}_{app}]^2} \label{e661}.
	\end{eqnarray}
	
	In this equation, $\overline{\Sigma}_{app}$ is the solution of the
	self-consistent Eq.~\eqref{e68c} at fixed $kT$, $K_c$, $R$, and $S_0$.
	
	If the correlator is in some sense a small quantity,  then due to the inequality in Eq.~\eqref{e148} the
	thermodynamics of the model system from
	Eq.~\eqref{e141} is well approximated by a system, having approximating
	Hamiltonian $H_{app}(U,\widetilde{X})$, which is the expression in Eq.~\eqref{AH1} with $X=\widetilde{X}$.
	Thus, the use of an approximating Hamiltonian is justified when the
	correlator is small enough. We will try to put these qualitative considerations
	on a more quantitative basis.
	
Let us consider the behavior of the correlator Eq.~\eqref{e661} at the extreme values of $K_s$ (0 and $\infty$).
	From Eq.~\eqref{e68d} it follows that when  $K_s \rightarrow 0$ at fixed $kT$,
	$K_c$, $R$, and $S_0$, then the correlator in Eq.~\eqref{e661} also tends to
	zero. From the same equation it follows that when $K_s \rightarrow \infty$,
	taking into account the result from Eq.~\eqref{bt4}  that $\overline{\Sigma}_{app}$ tends to a limited
	value, the correlator also tends to $\infty$.
	
	In order to estimate the correlator in Eq.~\eqref{e661}, measured in units $kT$, we shall  use the following numerical values of the quantities:\\
	$K_s\sim100$~erg/cm$^2$;\\
	$K_c\sim10^{-12}$erg;\\
	$R\sim10^{-3}$cm;\\
	$S_0\sim 4\pi R^2\sim 1.256\times 10^{-5}$cm$^2$;\\
	$\sigma_1=4\times 10^9$erg$^{-0.5}$.\\
	These values are typical for the experiments where analysis of the shape
	fluctuations of nearly spherical lipid vesicles were carried out \cite{Mel97}.
	With these values the quantity $\langle [{\cal A}(U)-\widetilde{X}]^2
	\rangle_{H_{app}(U,\widetilde{X})}$ from Eq.~\eqref{e661} is estimated as follows:
	\begin{eqnarray}
	\langle [{\cal A}(U) & - & \widetilde{X}]^2 \rangle_{H_{app}(U,\widetilde{X})}
	\nonumber \\
	&& = 8 \times 10^{-10} \sum_{n=2}^{n_{max}}
	\frac{2n+1}{[n(n+1)+\overline{\Sigma}_{app}]^2}.
	\label{167}
	\end{eqnarray}

	The value of the correlator is a measure of the error in
	the determination of the free energy $f[H_{app}(U,\widetilde{X})]$ in
	Eq.~\eqref{e62} with $X=\widetilde{X}$. We assume that it is small
	enough when the following inequality is fulfilled:
	\begin{equation}
	\frac{\langle [{\cal A}(U)-\widetilde{X}]^2
		\rangle_{H_{app}(U,\widetilde{X})}}{f[H_{app}(U,\widetilde{X})]} \ll 1
\label{IN1}
	\end{equation}
	Our numerical calculations showed that for the values of the quantities $K_s,K_c,R, S_0$ and $\sigma_1$ given above, the inequality in Eq.~\eqref{IN1} is valid. Consequently, the mean-field approximation, applied to the studied system, shows very good results.
	
	\section{Results and conclusions}
	
In the   Milner and Safran   phenomenological theory \cite{Mil87,Sei95},  the  "effective tension" $\overline{\sigma}$  is a free parameter which simulates an area constraint
on the fluctuation amplitudes in Eq.~\eqref{h21}. The mean square amplitudes  are used to determine experimentally the bending elasticity $K_c$ \cite{Fauc89}. The question is whether it is possible to include the experimental  determination of the stretching elasticity $K_s$ in this scenario? Having this in maind it is necessary to take into consideration the area dilation energy in the Hamiltonian of the fluctuating system.
	Usually,  in the simplest
	Hamiltonian approach the membrane area constraint is is guaranteed by a Lagrange
	multiplier $\sigma$ conjugated to the real area $S(U)$ \cite{Sei95,Bif10,F11,Hnf16}. This Lagrange  multiplier fixes the mean area $\langle
	S(U)\rangle_{H(U)}$ of the vesicle and is known as the "intrinsic tension"\cite{F11} or "internal tension"\cite{Hnf16}.
Its value cannot be directly experimentally measured. Its relation with the multiple other definitions
of membrane surface tension is  a matter of a longstanding debate (see e.g.\cite{Bif10,F11,Hnf16} and refs. therein).

Another approach for description of the fluctuating system is to present the  constraint $S(U)=const$ exactly by a $\delta$-function \cite{Sei95,Fap01}. It must be pointed out that
it is a quite  subtle  to study the thermodynamics of the system in the later case.  This requires to use the integral representation of the $\delta-$function  in the partition function.
 The corresponding calculations can be evaluated  using the method of steepest descend in the corresponding integral and are exactly valid in the thermodynamic limit \cite{Sei95,Fap01,Fp03}.

The methods,
	involving Lagrange multipliers allow the easier analytical
	of calculation the partition function for  fluctuating vesicles  with a
	fixed area,(see also refs.~\cite{Mil87,Sei95,Mel97}).
	However, this means to change the statistical ensemble.
	
	Although the two approaches
model two different physical situations, it is widely assumed
that the ensembles give equivalent results in the
thermodynamic limit, i.e., in the limit of infinitely large membranes \cite{F11,Hnf16}.
It is questionable  whether both approaches are totally equivalent.
The problem resembles the well-known problem for the
	equivalence of spherical and of mean spherical models of ferromagnetism
	which belong to different ensembles. It is well known that these two ensembles are,
	in general, statistically inequivalent (for a  discussion and list of references
	see Chapter 3 of the monograph \cite {Bdt00}). In the membrane fluctuation theories the above mentioned equivalence problem should be carefully
	reconsidered. In addition there is one more profound obstacle, namely infinitely large membrane at finite tension
	is an  object not at thermodynamic equilibrium \cite{S13} and the description in different
	ensembles should not give equivalent results.
	
	Our considerations avoids these problems being in accordance with the
	prescriptions of the statistical mechanics of finite-size systems.
The considered Hamiltonian $H(U)$ is nonlinear with respect to the squares
	$(u_n^m)^2$ of the amplitudes $u_n^m$, due to the nonlinearity of $H_s(U)$. This causes certain mathematical problems. To avoid this
	obstacle, one can follow the common approach based on the Habbard-Stratonovich transformation,
with the subsequent use of the saddle-point point approximation
	\cite{Fap01,Hi04,Lli11}. It turns out that
	the problem
	is  exactly solvable (only) in the thermodynamic limit
	\cite{Fap01,Hi04,Lli11,F11}. Let us recall that this aspect of the theory was
	already discussed in the context of  the spherical model of phase transitions
	in 1976 \cite{Bgp76}.
	
In the present paper we linearize the Hamiltonian in Eq.~\eqref{Hs1} using different approach. It is
	based on the Bogoliubov  variational inequalities. In our opinion this yields   a more clear picture of the proposed approximation. Moreover, the approximation is not related with
the notion of the thermodynamic limit.

In our approach the problem is reduced to solving a self-consistent equation  for the auxiliary variable $X$.
	For $X=\widetilde{X}$ this equation has a simple physical
	interpretation if it is presented in the form:
	\begin{eqnarray}
	&& \Bigg[\frac{1}{2}
	 \sum_{n=2}^{n_{max}}\sum_{m=-n}^n
	(n-1)(n+2) \langle(u_n^m)^2\rangle_{H_{app}(U,\widetilde{X})} \nonumber\\
	&& + 4\pi\Bigg] R^2  = A(\widetilde{X}),\label{fa}
	\end{eqnarray}
	where
	\begin{equation}
	A(\widetilde{X})= 4\pi R^2 + \sqrt{\frac{2S_0}{K_s}}\widetilde{X}.
	\end{equation}
	Comparing with the approach where the microscopic area  is fixed in the partition function through
 delta-function
(see, e.g., Eq. (47) in \cite{Sei95} and Eq. (15) in \cite{Fap01}), we see
	that the equality \eqref{fa} (valid for membrane parameters: $S_0,K_s, K_c, R$, and temperature $T$) imposes a "soft"  constraint on the  amplitudes of the shape fluctuations of the vesicle.
	It assures that the mean area  of the membrane (lhs of Eq.~\eqref{fa}) is equal to the area $A(\widetilde{X})$ (rhs of Eq.~\eqref{fa}).
If $\overline{\Sigma}_{app}(\widetilde{X})$ is considered as a fitting parameter, then the expressions from Eq.~\eqref{e67} for  $\langle (u_n^m)^2\rangle_{H_{app}(U,X)}$ can be used to determine  not only the bending elasticity $K_c$, but also the stretching elasticity  $K_s$, by using the measurable shape fluctuations of quasi-spherical vesicles.  One possible method for carrying out  such measurements is the treatment of the images in order to extract the vesicles' contours by means of phase contrast
	microscopy combined with fast image processing \cite{Fauc89}.

	In the present paper we showed that  theory $\grave{a}~la$  Milner and Safran can be
	deduced by an approach based on the Bogolyubov inequalities and the
	approximating Hamiltonian method which is in accordance to the principles of
	statistical mechanics. Our consideration reveals the principal possibility to extract the value of the
	stretching elasticity modulus $K_s$ of the vesicle membrane from the analysis of the
	shape fluctuations of nearly spherical lipid vesicles. It was proved that there is an interval of values of the
	vesicle membrane tension, for which the application of this method give precise
	enough results. Quite unexpectedly this approach is applicable
	when the stretching elasticity $K_s$ of the membrane tends to zero.

\section{Acknowledgements}
The work is partly supported by the Project for cooperation (2016/2018) between the JINR (Dubna) - ISSP-BAS (Sofia) and the Project with the Ministry of Education and Research, Bulgaria (National Science Fund, Grant DN 08-02/2016).
	
	\bibliography{Bivas_Tonchev_paper}
	
\end{document}